\documentclass[12pt]{article}
\usepackage[a4paper, top=16mm, text={170mm, 248mm}, includehead, includefoot, hmarginratio=1:1, heightrounded]{geometry}
\usepackage{amsmath,amssymb,mathrsfs,amsthm,tikz,shuffle,paralist}
\usepackage{dsfont}
\usepackage{color}
\definecolor{darkred}{RGB}{173,34,48}

\usetikzlibrary{snakes}
\usetikzlibrary{calc}
\usetikzlibrary{decorations}

\usepackage[all]{xypic}
\usepackage{jheppub}
\usepackage{hyperref}
\usepackage{subfigure}
\usepackage{caption}

\DeclareMathOperator{\Li}{Li} 

\title{Comments on all-loop constraints for scattering amplitudes and Feynman integrals}

\date{\today}
\author[a,b,c,d,e]{Song He}
\author[a,d]{Zhenjie Li} 
 \author[a,d]{Qinglin Yang}%
\affiliation[a]{CAS Key Laboratory of Theoretical Physics, Institute of Theoretical Physics, Chinese Academy of Sciences, Beijing 100190, China}
\affiliation[b]{
School of Fundamental Physics and Mathematical Sciences, Hangzhou Institute for Advanced Study, UCAS, Hangzhou 310024, China}
\affiliation[c]{ICTP-AP
International Centre for Theoretical Physics Asia-Pacific, Beijing/Hangzhou, China}
\affiliation[d]{School of Physical Sciences, University of Chinese Academy of Sciences, No.19A Yuquan Road, Beijing 100049, China}
\affiliation[e]{Peng Huanwu Center for Fundamental Theory, Hefei, Anhui 230026, P. R. China}
\emailAdd{songhe@itp.ac.cn}
\emailAdd{lizhenjie@itp.ac.cn}
\emailAdd{yangqinglin@itp.ac.cn}

\abstract{We comment on the status of  ``Steinmann-like" constraints, {\it i.e.} all-loop constraints on consecutive entries of the symbol of scattering amplitudes and Feynman integrals in planar ${\cal N}=4$ super-Yang-Mills, which have been crucial for the recent progress of the bootstrap program. Based on physical discontinuities and Steinmann relations, we first summarize all possible double discontinuities (or first-two-entries) for (the symbol of) amplitudes and integrals in terms of dilogarithms, generalizing well-known results for $n=6,7$ to all multiplicities. As our main result, we find that extended-Steinmann relations hold for all finite integrals that we have checked, including various ladder integrals, generic double-pentagon integrals, as well as finite components of two-loop NMHV amplitudes for any $n$; with suitable normalization such as minimal subtraction, they hold for $n=8$ MHV amplitudes at three loops. We find interesting cancellation between contributions from rational and algebraic letters, and for the former we have also tested cluster-adjacency conditions using the so-called Sklyanin brackets. Finally, we propose a list of possible last-two-entries for $n$-point MHV amplitudes derived from $\bar{Q}$ equations, which can be used to reduce the space of functions for higher-point MHV amplitudes. 
}

\begin{document}

\maketitle
\section{Introduction}
Recent years have witnessed enormous progress in computing and understanding analytic structures of scattering amplitudes in QFT. These developments have greatly pushed the frontier of perturbative calculations relevant for high energy experiments, and often they offer deep insights into the theory itself and exhibit surprising connections with mathematics. An outstanding example is the ${\cal N}=4$ super-Yang-Mills theory (SYM), where one can perform calculations that were unimaginable before and discover rich mathematical structures underlying them. For example, positive Grassmannian~\cite{Arkani-Hamed:2016byb} and the amplituhedron~\cite{Arkani-Hamed:2013jha} have provided a new geometric formulation for its planar integrand to all loop orders.

On the other hand, the modern bootstrap program in perturbative QFT aims at computing scattering amplitudes and other physical quantities by directly imposing analytic structure and physical constraints, without ever needing Feynman diagrams or loop integrands. Again the perfect laboratory where very impressive perturbative orders have been achieved is the planar ${\cal N}=4$ SYM amplitudes with $n=6,7$~\cite{Dixon:2011pw,Dixon:2014xca,Dixon:2014iba,Drummond:2014ffa,Dixon:2015iva,Caron-Huot:2016owq,Dixon:2016nkn,Drummond:2018caf, Caron-Huot:2019vjl, Caron-Huot:2019bsq, Dixon:2020cnr, Golden:2021ggj} (see \cite{Caron-Huot:2020bkp} for a review). The key of the bootstrap program is the construction of a function space where the amplitude lives: for $n=6,7$ (or more generally MHV and NMHV amplitudes with any $n$) conjecturally it is the space of generalized polylogarithm functions, which depend on $3(n{-}5)$ dual conformal invariant (DCI) cross-ratios, after subtracting infrared divergences {\it e.g.} by normalizing with Bern-Dixon-Smirnov (BDS) ansatz~\cite{Bern:2005iz}. As another deep connection between amplitudes and mathematics, it has been realized in~\cite{Golden:2013xva} that cluster algebras of Grassmannian $G(4,n)$ are directly relevant for the singularities of $n$-particle amplitudes, whose kinematics can be parametrized by momentum twistor~\cite{Hodges:2009hk}, or the space of $G(4,n)$ mod torus action. More precisely, the so-called ${\cal A}$-coordinates of $G(4,n)$ cluster algebras are related to symbol~\cite{Goncharov:2010jf, Duhr:2011zq} letters of amplitudes: the $9$ letters of six-particle amplitudes and $42$ letters of seven-particle ones are nicely explained by $A_3$ and $E_6$ cluster algebras, respectively.

Apart from the overall symbol alphabet, more refined restrictions on scattering amplitudes and Feynman integrals have proved to be crucial for the bootstrap program to higher orders for $n=6,7$. The most basic ones include the physical discontinuity~\cite{Gaiotto:2011dt} and the double discontinuity structure derived from the Steinmann relations~\cite{Steinmann1960a,Steinmann1960b}, which have reduced possible first-two-entries to only $6$ and $28$ weight-two symbols for $n=6,7$, respectively. Remarkably, the conjectural extended Steinmann (ES) relations, which basically extend Steinmann relations to higher multiple discontinuities, have been observed to hold for amplitudes and integrals for $n=6,7$\cite{Caron-Huot:2016owq,Dixon:2016nkn, Caron-Huot:2019bsq}. These ES relations impose similar constraints on any two consecutive entries of the symbol as Steinmann relations on the first two, which have been explicitly checked for $n=6$ (double-)pentagon ladders to arbitrary loops (and some for $n=7$) as well as $n=6$ amplitudes (up to $L=7$) and $n=7$ amplitudes (up to $L=4$), which are normalized by BDS-like subtraction~\cite{Alday:2009dv} and two-loop $n=8$ MHV amplitude with minimal subtraction~\cite{Golden:2018gtk}~\footnote{Interesting constraints similar to extended Steinmann relations have been found for Feynman integrals~\cite{Bourjaily:2020wvq}, form factors~\cite{Dixon:2020bbt} and even the wave-function of the universe~\cite{Benincasa:2020aoj}.}. In these cases, the ES relations can be alternatively formulated as cluster adjacency conditions~\cite{Drummond:2017ssj,Drummond:2018caf,Drummond:2018dfd}: only ${\cal A}$ coordinates that belong to the same cluster (for $A_3$ and $E_6$ for $n=6,7$ respectively) can appear adjacent in the symbol. As far as we know, ES relations/cluster adjacency has remained (mysterious) conjectures, and it is natural to ask if they extend to amplitudes and integrals for $n\geq 8$. It is in particular highly desirable to apply it to finite Feynman integrals to relatively high orders, as well as finite components of amplitudes, for higher multiplicities; among other things, such relations can provide powerful constraints on the function space for higher $n$ just as what they have achieved for $n=6,7$. This is the main question we hope to address in the paper. 

Beyond $n=6,7$, Grassmannian cluster algebras for $G(4,n)$ for $n\geq 8$ become infinite, and a certain truncation is needed to obtain a finite symbol alphabet. As already seen for one-loop N${}^2$MHV, amplitudes with $n\geq 8$ generally involve letters that cannot be expressed as rational functions of Pl\"ucker coordinates of the kinematics $G(4,n)/T$; more non-trivial algebraic letters appear in computations based on ${\bar Q}$ equations~\cite{CaronHuot:2011kk} for two-loop NMHV amplitudes for $n=8$ and $n=9$~\cite{Zhang:2019vnm, He:2020vob}, requiring extension of Grassmannian cluster algebras to include algebraic letters. Solutions to both problems has been proposed using tropical positive Grassmannian~\cite{speyer2005tropical} and related tools for $n=8$~\cite{Drummond:2019qjk, Drummond:2019cxm, Henke:2019hve, Arkani-Hamed:2019rds,Arkani-Hamed:2020cig, Herderschee:2021dez} and $n=9$~\cite{Henke:2021avn,Ren:2021ztg}
, as well as using Yangian invariants or the associated collections of plabic graphs~\cite{Mago:2020kmp, He:2020uhb,Mago:2020nuv,Mago:2021luw}.

On the other hand, ${\cal N}=4$ SYM has been an extremely fruitful laboratory for the study of Feynman integrals ({\it c.f.} \cite{ArkaniHamed:2010gh,Drummond:2010cz, Caron-Huot:2018dsv, Henn:2018cdp, Bourjaily:2018aeq, Herrmann:2019upk} and references therein).The connections to cluster algebras extend to individual Feynman integrals as well, {\it e.g.} the same $A_3$ and $E_6$ control $n=6,7$ multi-loop integrals in ${\cal N}=4$ SYM~\cite{Caron-Huot:2018dsv, Drummond:2017ssj}. Cluster algebra structures have also been discovered for Feynman integrals beyond those in planar ${\cal N}=4$ SYM~\cite{Chicherin:2020umh} including a five-particle alphabet which has played an important role in recent two-loop computations~\cite{Abreu:2018aqd,Chicherin:2018old,Chicherin:2018yne}. The knowledge of alphabet and more refined information can be used for bootstrapping Feynman integrals~\cite{Chicherin:2017dob, Henn:2018cdp} (see also \cite{Dixon:2020bbt}). In~\cite{He:2021esx,He:2021non}, we have identified (truncated) cluster algebras for the alphabets of a class of finite, dual conformal invariant (DCI)~\cite{Drummond:2006rz,Drummond:2007aua} Feynman integrals to high loops, based on recently-proposed Wilson-loop $d\log$ representation~\cite{He:2020uxy,He:2020lcu}. For ladder integrals with possible ``chiral pentagons" on one or both ends (without any square roots), we find a sequence of cluster algebras $D_2, D_3, \cdots, D_6$  for their alphabets, depending on $n$ and the kinematic configurations; for cases with square root, such as the $n=8$ double-penta ladder integrals, we find a truncated affine $D_4$ cluster algebra which (minimally) contain $25$ rational letters and $5$ algebraic ones. 

In this paper, we would like to use this rich collection of data for a systematic check of extended Steinmann relations beyond $n=6,7$. Before proceeding, let us briefly list all finite integrals we use as data in this paper. As the main example, we consider $n=8$ double-penta ladder integral, $I^{(L)}_{\text{dp}}(1,4,5,8)$ with four massless corners labelled by $1,4,5,8$, up to $L=4$ (whose alphabet contains $25+5$ letters mentioned above). Its collinear limit $8 \to 7$ gives $n=7$ $I^{(L)}_\text{dp}(1,4,5,7)$ with an alphabet of $D_4$ cluster algebra ($16$ rational letters). Simpler examples include penta-box ladder $I^{(L)}_\text{pb}$ up to $L=4$ (alphabet is $D_3$ cluster algebra) and the well-known box ladder $I^{(L)}_\text{b}$ to all loops (alphabet is $D_2 \sim A_1^2$ cluster algebra). 
\begin{figure}[htbp]
    \begin{tikzpicture}[baseline={([yshift=-.5ex]current bounding box.center)},scale=0.15]
                \draw[black,thick] (0,5)--(-5,5)--(-5,0)--(0,0)--cycle;
                \draw[black,thick] (1.93,5.52)--(0,5)--(0.52,6.93);
                \draw[black,thick] (1.93,-0.52)--(0,0)--(0.52,-1.93);
                \draw[thick,densely dashed] (-10,0) -- (-5,0);
                \draw[thick,densely dashed] (-10,5) -- (-5,5);
                \draw[black,thick] (-10,0)--(-10,5)--(-15,5)--(-15,0)--cycle;
                \draw[black,thick] (-16.93,5.52)--(-15,5)--(-15.52,6.93);
                \draw[black,thick] (-16.93,-0.52)--(-15,0)--(-15.52,-1.93);
                \filldraw[black] (1.93,6) node[anchor=west] {{$2$}};
                \filldraw[black] (0.52,6.93) node[anchor=south] {{$1$}};
                \filldraw[black] (1.93,-1) node[anchor=west] {{$3$}};
                \filldraw[black] (0.52,-1.93) node[anchor=north] {{$4$}};
                \filldraw[black] (-16.93,6) node[anchor=east] {{$7$}};
                \filldraw[black] (-15.52,6.93) node[anchor=south] {{$8$}};
                \filldraw[black] (-16.93,-1) node[anchor=east] {{$6$}};
                \filldraw[black] (-15.52,-1.93) node[anchor=north] {{$5$}};
                \filldraw[black] (-7.52,6.93) node[anchor=south] {{$x_1$}};
                \filldraw[black] (1.93,2.02) node[anchor=west] {{$x_3$}};
                \filldraw[black] (-7.52,-1.93) node[anchor=north] {{$x_5$}};
                \filldraw[black] (-16.93,2.02) node[anchor=east] {{$x_7$}};
            \end{tikzpicture}
\begin{tikzpicture}[baseline={([yshift=-.5ex]current bounding box.center)},scale=0.15]
                \draw[black,thick] (0,0)--(0,5)--(4.76,6.55)--(7.69,2.5)--(4.76,-1.55)--cycle;
                \draw[decorate, decoration=snake, segment length=12pt, segment amplitude=1.5pt, black,thick] (4.76,6.55)--(4.76,-1.55);
                \draw[black,thick] (9.43,1.5)--(7.69,2.5)--(9.43,3.5);
                \draw[black,thick] (4.76,6.55)--(5.37,8.45);
                \draw[black,thick] (4.76,-1.55)--(5.37,-3.45);
                \draw[black,thick] (0,5)--(-5,5)--(-5,0)--(0,0);
                \draw[thick,densely dashed] (-10,0) -- (-5,0);
                \draw[thick,densely dashed] (-10,5) -- (-5,5);
                \draw[black,thick] (-10,0)--(-10,5)--(-15,5)--(-15,0)--cycle;
                \draw[black,thick] (-16.93,5.52)--(-15,5)--(-15.52,6.93);
                \draw[black,thick] (-16.93,-0.52)--(-15,0)--(-15.52,-1.93);
                \filldraw[black] (5.37,8.45) node[anchor=south] {{$1$}};
                \filldraw[black] (5.37,-3.45) node[anchor=north] {{$4$}};
                \filldraw[black] (9.43,1.4) node[anchor=west] {{$3$}};
                \filldraw[black] (9.43,3.6) node[anchor=west] {{$2$}};
                \filldraw[black] (-16.93,5.52) node[anchor=east] {{$7$}};
                \filldraw[black] (-15.52,6.93) node[anchor=south] {{$8$}};
                \filldraw[black] (-16.93,-0.52) node[anchor=east] {{$6$}};
                \filldraw[black] (-15.52,-1.93) node[anchor=north] {{$5$}};
            \end{tikzpicture}
\begin{tikzpicture}[baseline={([yshift=-.5ex]current bounding box.center)},scale=0.15]
        \draw[black,thick] (0,0)--(0,5)--(4.76,6.55)--(7.69,2.5)--(4.76,-1.55)--cycle;
        \draw[black,thick] (-15,5)--(-19.76,6.55)--(-22.69,2.5)--(-19.76,-1.55)--(-15,0);
        \draw[decorate, decoration=snake, segment length=12pt, segment amplitude=2pt, black,thick] (4.76,6.55)--(4.76,-1.55);
        \draw[decorate, decoration=snake, segment length=12pt, segment amplitude=2pt, black,thick] (-19.76,6.55)--(-19.76,-1.55);
        \draw[black,thick] (9.43,1.5)--(7.69,2.5)--(9.43,3.5);
        \draw[black,thick] (4.76,6.55)--(5.37,8.45);
        \draw[black,thick] (4.76,-1.55)--(5.37,-3.45);
        \draw[black,thick] (0,5)--(-5,5)--(-5,0)--(0,0);
        \draw[black,thick,densely dashed] (-5,5)--(-10,5);
        \draw[black,thick,densely dashed] (-5,0)--(-10,0);
        \draw[black,thick] (-10,0)--(-10,5)--(-15,5)--(-15,0)--cycle;
        \draw[black,thick] (-19.76,6.55)--(-20.37,8.45);
        \draw[black,thick] (-19.76,-1.55)--(-20.37,-3.45);
        \draw[black,thick] (-24.69,3.5)--(-22.69,2.5)--(-24.69,1.5);
        \filldraw[black] (-20.37,8.45) node[anchor=south] {{8}};
        \filldraw[black] (5.37,8.45) node[anchor=south] {{1}};
        \filldraw[black] (9.43,3.5) node[anchor=west] {{2}};
        \filldraw[black] (9.43,1.5) node[anchor=west] {{3}};
        \filldraw[black] (5.37,-3.45) node[anchor=north] {{4}};
        \filldraw[black] (-20.37,-3.45) node[anchor=north] {{5}};
        \filldraw[black] (-24.69,1.5) node[anchor=east] {{6}};
        \filldraw[black] (-24.69,3.5) node[anchor=east] {{7}};
    \end{tikzpicture}
    \caption{The box ladder $I_\text{b}$,  penta-box ladder $I_\text{pb}$ and  double-penta ladder $I_\text{dp}(1,4,5,8)$ for $n=8$.}\label{fig1}
\end{figure}
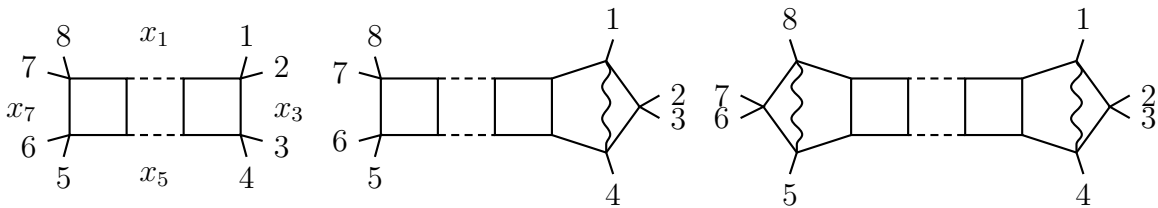

Besides these ladder integrals up to $n=8$, we will also use the most general two-loop double-pentagon integrals computed in~\cite{He:2020lcu}, which we denote as $\mathcal{I}_{\text{dp}}(i,j,k,l)$. These integrals become generic at $n=12$ {\it e.g.} $(i,j,k,l)=(1,4,7,10)$ (depending on $13$ variables) whose alphabet has $164$ rational letters and $5$ multiplicative independent algebraic letters for each of the $16$ distinct square roots. For $n=8$, we have {\it e.g.} $\mathcal{I}_{\text{dp}}(1,3,5,7)$ (depending on $9$ variables) whose alphabet has $100$ rational letters and $2\times 5$ algebraic ones (with two different square roots).
\begin{align*}
\mathcal{I}_{\mathrm{dp}}(i,j,k,l) \,\, = \,\,
\begin{tikzpicture}[baseline={([yshift=-.5ex]current bounding box.center)},scale=0.16]
\draw[black,thick](0,0)--(0,5)--(4.75,6.54)--(7.69,2.50)--(4.75,-1.54)--cycle;
\draw[black,thick](0,5)--(-4.75,6.54)--(-7.69,2.50)--(-4.75,-1.54)--(0,0);
\draw[decorate, decoration=snake, segment length=12pt,segment amplitude=2pt,black,thick] (4.75,6.54)--(4.75,-1.54);
\draw[decorate, decoration=snake, segment length=12pt,segment amplitude=2pt,black,thick] (-4.75,6.54)--(-4.75,-1.54);
\draw[black,thick](-0.9,6.5)--(0,5)--(0.9,6.5);
\filldraw[black]  (0,6) circle [radius=1.5pt];
\filldraw[black]  (-0.25,6) circle [radius=1.5pt];
\filldraw[black]  (0.25,6) circle [radius=1.5pt];
\draw[black,thick](-0.9,-1.5)--(0,0)--(0.9,-1.5);
\filldraw[black]  (0,-1) circle [radius=1.5pt];
\filldraw[black]  (-0.25,-1) circle [radius=1.5pt];
\filldraw[black]  (0.25,-1) circle [radius=1.5pt];
\draw[black,thick](4.75,6.54)--(6,7.94);
\draw[black,thick](4.75,-1.54)--(6,-3.04);
\draw[black,thick](9.19,1.6)--(7.69,2.50)--(9.19,3.4);
\draw[black,thick](-4.75,6.54)--(-6,7.94);
\draw[black,thick](-4.75,-1.54)--(-6,-3.04);
\draw[black,thick](-9.19,1.6)--(-7.69,2.50)--(-9.19,3.4);
\filldraw[black]  (9,2.5) circle [radius=1.5pt];
\filldraw[black]  (9,2.25) circle [radius=1.5pt];
\filldraw[black]  (9,2.75) circle [radius=1.5pt];
\filldraw[black]  (-9,2.5) circle [radius=1.5pt];
\filldraw[black]  (-9,2.25) circle [radius=1.5pt];
\filldraw[black]  (-9,2.75) circle [radius=1.5pt];
\filldraw[black] (6,7.94) node[anchor=south west] {{$k$}};
\filldraw[black] (6,-3.04) node[anchor=north west] {{$l$}};
\filldraw[black] (-6,7.94) node[anchor=south east] {{$j$}};
\filldraw[black] (-6,-3.04) node[anchor=north east] {{$i$}};
\end{tikzpicture}
\end{align*}
Remarkably, these integrals directly give IR finite components of two-loop NMHV amplitudes to all multiplicities! More precisely, for non-adjacent $i,j,k,l$ the component $\chi_i \chi_j \chi_k \chi_l$ of NMHV amplitude (after stripping off MHV tree)/Wilson loop is simply given by ${\cal I}_{\rm dp} (i,j,k,l)-{\cal I}_{\rm dp} (j,k,l,i)$; they represent the simplest class of NMHV components (while most generic for large $n$), which are also free of square roots~\cite{Bourjaily:2019igt} (as first computed for $n=8$ and for higher $n$ using $\bar{Q}$ method~\cite{Zhang:2019vnm,He:2020vob}). 

In section 3, we will explicitly check ES relations and find positive answer for all these finite Feynman integrals (up to $L=4$) and two-loop NMHV (finite) amplitudes with $n\geq 8$; note that the latter provides all-multiplicity evidence for ES relations for finite component amplitudes without the need of any subtraction/normalization. We will also confirm ES relations using the three-loop $n=8$ MHV amplitudes obtained very recently using ${\bar Q}$ equations by one of the authors and Chi Zhang~\cite{Li2021}. Similar to the BDS-like subtraction for $n=6,7$ case, we need such normalization for higher-point MHV amplitudes and in particular for the case when $n$ is a multiple of $4$, such as $n=8$, the BDS-like subtraction is not defined and we need {\it e.g.} minimal subtraction. Note that for two-loop MHV amplitudes, cluster adjacency, which is closely related to ES relations, has been checked in~\cite{Golden:2019kks} using Sklyanin bracket, which also requires minimal subtraction for $n=8$. It is not at all obvious to us that with the same subtraction, three-loop $n=8$ MHV amplitude also satisfies ES relations! 

In all these cases except for $I_{\rm pb}$ and $I_{\rm dp} (1,4,5,7)$, we need to take into account algebraic letters, and while most ES relations directly hold for the rational part, naively there is a violation of ES relations in this part for remaining pairs which appear in cross-ratio of four-mass kinematics; it is remarkable that by including the algebraic part which also depends on such four-mass kinematics, the violation is cancelled and we have ES relations for the full answer.
Note that while ES relations can be imposed without explicitly referring to the alphabet, it is more subtle to extend cluster adjacency to higher $n$ whose alphabet also contains algebraic letters; for rational ones, the so-called Sklyanin bracket~\cite{Sklyanin:1983ig} can be used, and we will see that for all those rational letters which do not violate ES relations, we can test that cluster adjacency is also satisfied.

In addition to ES relations which apply to all consecutive entries of the symbol, it is well known that the first two entries in general and last two entries for MHV amplitudes are not the same as before. As we have mentioned, given the physical discontinuity or the first entry condition, Steinmann-satisfying first two entries are much more constrained than subsequent pairs. In section 2, we will list all such (integrable) first-two-entries, which amounts to having the weight-$2$ space of Steinmann-satisfying dilogarithm functions. We list all such weight-$2$ functions for any $n$, in terms of ${\rm Li}_2$ and $\log \log$. As extracted from the box functions, entries of the $\Li_2$ part will automatically DCI, while entries of $\log\log$ functions are simply $x_{ij}^2$. We will give the general counting of these first-two-entries to all $n$, and check this ansatz by all the data we have, including amplitudes and finite integrals.

In section 4 which has a different flavor than the rest of the paper, we will also provide a list of all possible last two entries for MHV amplitudes. The reason we study them is that last entries of MHV amplitudes have been known since~\cite{Caron-Huot:2011dec} which can be derived from ${\bar Q}$ equations and have played an important role for the bootstrap program. It is known again for $n=6,7$ that similar constraints can be derived for last two entries, again by using ${\bar Q}$ equations. We will explicitly go through the situations for $n\leq9$ and get all possible $\mathcal{A}$-coordinates pairs on the last-two-entries. Furthermore, for $n=7$ we reorganize the pairs into manifestly DCI integrable symbols, showing that there are $139$ allowed combinations. We will see that numbers of the last-two-entries from $\bar Q$ equations remain very limited as $n$ grows, leaving stronger constraints on the function space of MHV amplitudes.

\section{First-two-entries from Steinmann relations}

As a warm-up exercise, let us first present our conjecture for all possible first-two-entries of integrals/amplitudes. The starting point is that physical discontinuities, or the first entries of the symbol, for any amplitudes/integrals in planar ${\cal N}=4$ SYM, must correspond to {\it planar variables} $x_{a,b}^2=0$; this is even true for any planar theory, and what is special about SYM is the dual conformal symmetry which forces the first entries to be DCI combinations. Note that out of all the $n(n{-}3)/2$ planar variables, which we write as $\langle a{-}1 a b{-}1 b \rangle $, $n$ of them are {\it frozen variables} ($\langle i i{+}1 i{+}2 i{+}3\rangle$ for $i=1, \cdots, n$), thus we have $n(n{-}5)/2$ unfrozen Pl\"ucker variables which correspond to diagonals of $n$-gon except for the $n$ ``shortest" ones. Since we are interested in variables that are unfrozen, these are all we can have for the first entry, and equivalently one can construct the same number of DCI combinations {\it e.g.} $v_{i,j}:=x_{i,j}^2 x_{i{+}2, j{+}2}^2/(x_{i, i+2}^2 x_{j, j{+}2}^2)$. For example, for $n=6,7$, these correspond to the well known $3$ and $7$ first entries, and for $n=8$ it is easy to see that we have $12$ first entries, which are given by $x^2_{i,i{+}3}$ for $i=1,\cdots, 8$ and $x^2_{i, i{+}4}$ for $i=1,\cdots, 4$ (or corresponding $v_{i, i{+}3}$ and $v_{i, i{+}4}$). 

What about the first-two-entries? The most important constraint on the first-two-entries is the Steinmann relations, which says that the double discontinuities taken in overlapping channels of any amplitude should vanish. A convenient way to proceed is to start at one loop, where all finite amplitudes/integrals can be expanded in terms of (finite part of) box integrals. First we note that the only finite box integrals are the four-mass ones; given four dual points $x_a, x_b, x_c, x_d$ with $a,b,c,d$ cyclically ordered and each adjacent pair at least separated by $2$, we have the (normalized) four-mass box integral 
\begin{gather}
  I_{a,b,c,d}(u,v):={\rm Li}_2(1-
  z)-{\rm Li_2}(1-\bar{z})+\tfrac{1}{2} \log \biggl(\frac {z}{\bar{z}}\biggr) \log (v) \\
\text{with } u=u_{a,b,c,d}=z\bar{z}\:,\:v=u_{b,c,d,a}=(1-z)(1-\bar{z})\,,
\end{gather}
where $u_{a,b,c,d}:=x_{a,b}^2 x_{c,d}^2/(x_{a,c}^2 x_{b,d}^2)$ and similarly for $u_{b,c,d,a}$. It is straightforward to check that $I_{a,b,c,d}$ satisfies Steinmann relations as we will see in the next section. We can also go to lower-mass cases but they suffer from infra-red (IR) divergences, and certain regularization is needed. One way to preserve DCI is to introduce the DCI-regulator as in~\cite{Bourjaily:2013mma}, but we find that the resulting finite part of the lower-mass box integrals violate Steinmann relations. More precisely, it is possible to write the ${\rm Li}_2$ part of lower-mass boxes in a way that respects Steinmann relations, but the $\log \log$ part does not, thus we will discuss these two parts separately for them. Our main conjecture is that for finite integrals and amplitudes to all loops, the first two entries that satisfy Steinmann relations can only be extracted from such box functions where the $\log \log$ part need to be treated separately. In addition to the four-mass box cases, we will encounter the following ${\rm Li}_2$ functions for three-mass and two-mass-easy boxes~\footnote{For two-mass-hard or one-mass boxes, only a constant ${\rm Li}_2(1)$ can appear in this part; as we are listing first-two-entries at the symbol level, we ignore such constant for now.}:
\begin{equation}
{\rm Li}_2 (1-1/u_{c,a{-}1, a, b}), \quad  {\rm Li}_2 (1-1/u_{c,a{-}1, a, c{-}1})
\end{equation}
where one can check that indeed Steinmann relations are satisfied: the second-entry that can enter are just $1-1/u_{c,a{-}1, a, b}$ and $1-1/u_{c,a{-}1, a, c{-}1}$. It is easy to count the total number of four-mass boxes and such ${\rm Li}_2$ functions for lower-mass cases: there are $n \choose 4$ box functions and $n(n{-}4)$ of them are two-mass-hard/one-mass boxes which only give ${\rm Li}_2(1)$, thus we have $n(n{-}5)(n^2-n-18)/24$ such functions. For example, for $n=6,7$, there are $3$ and $14$ ${\rm Li_2}$ functions, which can be chosen to be ${\rm Li}_2(1-a_i)$ for $i=1,2,3$ and ${\rm Li}_2(1-a_{1,i})$ for $i=1, \cdots, 7$, respectively~\cite{Dixon:2016nkn}. Now for $n=8$, we have $2$ four-mass boxes, and $36$ ${\rm Li}_2$ functions from lower-mass boxes, which we can choose to be $\operatorname{Li}_2(1-1/u)$ with $u$ in
\begin{align*}
\bigl\{&u_{1,3,4,6},u_{1,3,4,7},u_{1,3,4,8},u_{1,4,5,7},u_{1,4,5,8},u_{1,5,6,8},u_{2,4,5,1},u_{2,4,5,7},u_{2,4,5,8},u_{2,5,6,1},\\
&u_{2,5,6,8},u_{2,6,7,1},u_{3,5,6,1},u_{3,5,6,2},u_{3,5,6,8},u_{3,6,7,1},u_{3,6,7,2},u_{3,7,8,2},u_{4,6,7,1},u_{4,6,7,2},\\
&u_{4,6,7,3},u_{4,7,8,2},u_{4,7,8,3},u_{5,7,8,2},u_{5,7,8,3},u_{5,7,8,4},u_{5,8,1,3},u_{6,1,2,4},u_{6,8,1,3},u_{6,8,1,4},\\
&u_{7,1,2,4},u_{7,1,2,5},u_{7,2,3,5},u_{8,2,3,5},u_{8,2,3,6},u_{8,3,4,6}\bigr\}.
\end{align*}

Having fixed all possible ${\rm Li}_2$ part, we move to the $\log \log$ part, and it is clear which of them satisfy Steinmann relations:
\begin{equation}
\log x_{i,j}^2 \log x_{k,l}^2\,,\quad {\rm } (i,j) \sim (k,l)
\end{equation}
where $(i,j) \sim (k,l)$ means that the two diagonals (neither are the shortest ones) do not cross each other. Naively these do not respect DCI, but as we have mentioned above such $x_{i,j}^2$ can be converted to cross-ratios (with the help of frozen ones), which need to be done carefully with respect to Steinmann relations. For example, for $n=6,7$, there are exactly $3$ and $14$ such $\log\log$ functions respectively: $\log^2 x_{i,i{+}3}^2$ with $i=1,2,3$ for $n=6$, and $\log^2 x_{i, i{+}3}^2$ and $\log x^2_{i, i{+}3} \log x^2_{i{+}3, i{+}6}$ with $i=1,\cdots, 7$ for $n=7$. For $n=8$, we find $40$ $\log(x)\log(y)$ functions for $(x,y)$ in 
\begin{align*}
\bigl\{
&(x_{1,4}^2 , x_{1,4}^2),\,
(x_{1,4}^2 , x_{1,5}^2),\,
(x_{1,4}^2 , x_{1,6}^2),\,
(x_{1,4}^2 , x_{4,8}^2),\,
(x_{1,4}^2 , x_{5,8}^2),\,
(x_{1,4}^2 , x_{4,7}^2),\,
(x_{1,5}^2 , x_{1,5}^2),\,\\
&(x_{1,5}^2 , x_{1,6}^2),\,
(x_{1,5}^2 , x_{5,8}^2),\,
(x_{1,5}^2 , x_{2,5}^2),\,
(x_{1,6}^2 , x_{1,6}^2),\,
(x_{1,6}^2 , x_{2,5}^2),\,
(x_{1,6}^2 , x_{2,6}^2),\,
(x_{1,6}^2 , x_{3,6}^2),\,\\
&(x_{3,8}^2 , x_{3,8}^2),\,
(x_{3,8}^2 , x_{4,8}^2),\,
(x_{3,8}^2 , x_{5,8}^2),\,
(x_{3,8}^2 , x_{3,6}^2),\,
(x_{3,8}^2 , x_{3,7}^2),\,
(x_{3,8}^2 , x_{4,7}^2),\,
(x_{4,8}^2 , x_{4,8}^2),\,\\
&(x_{4,8}^2 , x_{5,8}^2),\,
(x_{4,8}^2 , x_{4,7}^2),\,
(x_{5,8}^2 , x_{5,8}^2),\,
(x_{5,8}^2 , x_{2,5}^2),\,
(x_{2,5}^2 , x_{2,5}^2),\,
(x_{2,5}^2 , x_{2,6}^2),\,
(x_{2,5}^2 , x_{2,7}^2),\,\\
&(x_{2,6}^2 , x_{2,6}^2),\,
(x_{2,6}^2 , x_{2,7}^2),\,
(x_{2,6}^2 , x_{3,6}^2),\,
(x_{2,7}^2 , x_{2,7}^2),\,
(x_{2,7}^2 , x_{3,6}^2),\,
(x_{2,7}^2 , x_{3,7}^2),\,
(x_{2,7}^2 , x_{4,7}^2),\,\\
&(x_{3,6}^2 , x_{3,6}^2),\,
(x_{3,6}^2 , x_{3,7}^2),\,
(x_{3,7}^2 , x_{3,7}^2),\,
(x_{3,7}^2 , x_{4,7}^2),\,
(x_{4,7}^2 , x_{4,7}^2)
\bigr\}.
\end{align*}

For general $n$, it is straightforward to count that there are $n(n{-}3)(n{-}4)(n{-}5)/12$ such $\log \log$ functions. In total, there are $n(n{-}5)(n^2-n+5)/8$ first-two-entries which can be derived from box functions and satisfy Steinmann relations, and we conjecture that they are all we need for finite integrals and Steinmann-respecting amplitudes. For example, we have checked that in all two-loop $n=8$ finite Feynman integrals in Figure \ref{fig1} and ${\cal I}_{\rm dp} (1,3,5,7)$, exactly these $78$ first-two-entries for $n=8$ can appear. Moreover, we have checked first-two entries of  minimally normalized of two-loop and three-loop MHV amplitudes for $n=8$, exactly all $78$ first-two-entries appear.

\section{Extended-Steinmann relations for integrals and amplitudes}

As discussed in the last section, the Steinmann relations that restrict the for amplitudes or Feynman integrals give constraints on the first-two entries. The extended Steinmann relations are just generalizing these constraints to any two consecutive entries, thus restricting iterated discontinuity at any depth. Let us first write these relations as follows: define a linear ``discontinuity'' map in the space of symbols
\begin{equation}
\operatorname{Disc}^k_{x=c} (a_1\otimes \cdots \otimes a_n):=\operatorname{Disc}_{x=c}(\log a_k)(a_1\otimes \cdots \otimes \widehat{ a_k}\otimes\cdots \otimes a_n),
\end{equation}
where we normalize the discontinuity map
that computes the monodromy around $x = c$ such that $\operatorname{Disc}_{x=c}\log (x-c) =1$.
Then the extended Steinmann relations for a amplitude or Feynman integral $F$ of weight $w$ are
\begin{equation}
\operatorname{Disc}^s_{x_{ij}^2=0}(\operatorname{Disc}^s_{x_{kl}^2=0}(\mathcal S(F)))=0
\end{equation}
for any $1\leq s\leq w-1$ and any two overlapping channels $x_{ij}^2$, $x_{kl}^2$, {\it i.e.} $(ij) \not\sim (kl)$. For $s=1$ this reduces to the Steinmann relations (the second $\operatorname{Disc}^1_{x=0}$ acting on the second-entry of the original symbol). Note that for $s\neq 1$, $\operatorname{Disc}^s_{x=c}$ of a  integrable symbol may not be integrable, so it's not well-defined in the space of functions.

If any two letters of $F$ do not share the same branch point, $\operatorname{Disc}_{x=0}(\log(y))=0$ for any two distinct letters $x$ and $y$, then $\operatorname{Disc}^k_{x=0}(F)$ is given by clipping the $k$-th entry $x$ off in the symbol of $F$ (after deleting all terms whose $k$-th entry is not $x$), and ES relations simply become constraints of adjacent letters in the symbol. This is indeed how ES relations are checked for BDS-like normalized amplitudes for $n=6,7$ (up to $L=7$ and $L=4$ respectively) as well as certain Feynman integrals (such as $n=6$ double-pentagon ladders). Note that this is always the case when the alphabet contains only rational letters, which cannot share any branch point since by definition they are multiplicative independent. For instance, we have also checked the double-penta ladder integrals $I_\text{dp}^{(L)}(1,4,5,7)$ with $n=7$ and the penta-box ladders $I_\text{pb}^{(L)}$ with $n=8$ up to $L=4$, and find that they indeed satisfy ES relations in exactly the same way as finite integrals for $n=6,7$.

\paragraph{Extended Steinmann relations with algebraic letters} However, generally amplitudes or Feynman integrals for $n\geq 8$ involve square roots, and algebraic letters may share the same branch point $x_{ij}^2=0$, such that extended Steinmann relations cannot be directly formulated as constraints on which letters can appear next to each other in the symbol. Let us illustrate this phenomenon with the simplest case: we consider the Steinmann relations of (normalized) four-mass box function $I_{2468}$:
\[
  \mathcal S(I_{2468})=\frac{1}{2} \biggl(u\otimes\frac{1-\bar z}{1-z}+v\otimes \frac{z}{\bar z}\biggr),
\]
where 
\begin{equation}\label{z2468}
u=\frac{x_{24}^2x_{68}^2}{x_{26}^2x_{48}^2}=z \bar z,\quad v=\frac{x_{46}^2x_{82}^2}{x_{26}^2x_{48}^2}=(1-z)(1-\bar z).
\end{equation}
The logarithms of these variables share 4 branch points $x_{24}^2,x_{68}^2,x_{26}^2,x_{48}^2=0$, and only $x_{26}^2,x_{48}^2=0$ are two overlapping channels (the two diagonals cross each other), and now we consider the Steinmann relation for this pair.

Near the branch point $x_{48}^2= 0$,
\[
\frac{1}{z\bar z}=u^{-1}=\frac{x_{26}^2x_{48}^2}{x_{24}^2x_{68}^2}\sim 0\quad \text{and}\quad 
\frac{1}{(1-z)(1-\bar z)}=v^{-1}=\frac{x_{26}^2x_{48}^2}{x_{46}^2x_{82}^2}\sim 0,
\]
so $z^{-1},(1- z)^{-1}\propto x_{48}^2$ or $\bar z^{-1},(1- \bar z)^{-1}\propto x_{48}^2$ depending on the region of kinematics. Here we choose the second one, and then
\[
\operatorname{Disc}_{x_{48}^2=0}\log\left(\frac{z}{\bar z}\right)=1,\quad \operatorname{Disc}_{x_{48}^2=0}\log\left(\frac{1-\bar z}{1-z}\right)=-1.
\]
Therefore, we see that algebraic letters can contribute to the discontinuity of a physical channel, and we can easily verify the Steinmann relation 
\[
\operatorname{Disc}^1_{x_{48}^2=0}\operatorname{Disc}^1_{x_{26}^2=0}\mathcal S(I_{2468})=\frac{1}{2}((-1)\cdot(-1)+(-1)\cdot 1)=0
\]
and similarly $\operatorname{Disc}^1_{x_{26}^2=0}\operatorname{Disc}^1_{x_{48}^2=0}\mathcal S(I_{2468})=0$.

A direct generalization of four-mass box is the  box-ladder $I_{b}^{(L)}$, which is the first example with square roots for checking extended Steinmann relations. The $L$-loop box-ladder function \cite{Usyukina:1993ch,Broadhurst:2010ds} has a well-known expression:
\begin{equation}\label{box-ladder}
I_{\text{b}}^{(L)}(\alpha,\bar\alpha)=\sum_{j=L}^{2 L} \frac{j ![-\log (\alpha \bar{\alpha})]^{2 L-j}}{L !(j-L) !(2 L-j) !}\left[\mathrm{Li}_{j}(\alpha)-\mathrm{Li}_{j}(\bar{\alpha})\right],
\end{equation}
where $\alpha$ and $\bar \alpha$ are defined by $
\alpha=\frac{z}{z-1}, \bar \alpha=\frac{\bar z}{\bar z-1}$. Note the alphabet of $\mathcal S(I_{\text{p}}^{(L)})$ for $L\geq 1$ is simply $D_2$:
\[
\{\alpha,\,1-\alpha,\,\bar \alpha,\,1-\bar\alpha\}, 
\]
and the only letter with non-zero discontinuity around the branch point $x_{48}^2=0$ or $x_{26}^2=0$ is $1-\alpha$ or $1-\bar\alpha$ depending on the region of kinematics. Thus we have seen the advantage of using $\alpha, \bar{\alpha}$ variables: the extended Steinmann relations for the box-ladder are simply the statement that there are no two consecutive $1-\alpha$ or $1-\bar \alpha$ in the symbol of $I_{\text{p}}^{(L)}(\alpha,\bar\alpha)$. This is quite clear from the expression eq.\eqref{box-ladder} that there is no two consecutive $1-\alpha$ in $\mathcal S(\mathrm{Li}_{j}(\alpha))=-(1-\alpha)\otimes \alpha\otimes \cdots \otimes \alpha$ and similarly for $\bar \alpha$ (obviously no contribution from powers of $\log (\alpha \bar \alpha)$). 

Next we move to the class of integrals with more non-trivial algebraic letters, namely double-pentagon ladder $I_{\text{dp}}^{(L)}(1,4,5,8)$, which have been computed up to $L=4$ with $25$ rational letters and $5$ independent algebraic letters~\cite{He:2021non}. First of all, since the algebraic letters only contain the square root as that of four-mass box $I_{2468}$, which only involve one pair of overlapping channels $x_{26}^2$ and $x_{48}^2$ as above. For any other pair of overlapping channels $x_{ij}^2$ and $x_{kl}^2$, the extended Steinmann relations simply mean that rational letters $\langle i{-}1\,i\,j{-}1\,j\rangle$ and $\langle k{-}1\,k,l{-}1\,l\rangle$ are not adjacent in the ``rational" part of the symbol of $I_{\text{dp}}^{(L)}(1,4,5,8)$. This is indeed the case as we have checked through $L=4$.

What we have found is that the only minor violation in the ``rational" part involves rational letters $\langle1256\rangle$ and $\langle3478\rangle$ in consecutive entries; in fact, they have only appeared next to each other at second and third entries. This is good news since the ``irrational" part only contains algebraic letters at these two entries~\cite{He:2021non}, which also contribute to double discontinuities involving this pair. Recall that the $5$ algebraic letters are of the form $(z-a)/(\bar z-a)$ for 
\[
a=0,\,\,1,\,\,\frac{\langle 1234\rangle \langle 1568\rangle}{\langle 1256\rangle \langle 1348\rangle},\,\,\frac{\langle 1234\rangle \langle 4578\rangle}{\langle 1245\rangle \langle 3478\rangle},\,\,
\biggl(1-\frac{\langle 1578\rangle \langle 3456\rangle}{\langle 1345\rangle \langle 5678\rangle}\biggr)^{-1}
\]
with $z$, $\bar z$ defined as in eq.\eqref{z2468}. Near the branch point $\langle 1256\rangle\propto x_{26}^2\sim 0$,  discontinuities of algebraic letters are given by 
\[
\operatorname{Disc}_{x_{26}^2=0} \log\biggl(\frac{z-a}{\bar z-a}\biggr)=
\begin{cases}
\pm 1, & \text{if $a^{-1}\not \propto x_{26}^2$},\\
0, &\text{if $a^{-1}\propto x_{26}^2$}
\end{cases},
\]
where the choices of $\pm 1$ depend on the kinematic region, but they should be the same on a given region. Similarly, we can calculate $\operatorname{Disc}_{x_{48}^2=0}\log$ of these algebraic letters. Remarkably, after taking into account the contribution from ``irrational" part, we find that the double discontinuity (at second and third entry) indeed vanishes, thus $I_{\text{dp}}^{(L)}(1,4,5,8)$ satisfy the extended Steinmann relations (at least up to $L=4$)!

So far we have focused on integrals up to $n=8$, but it is straightforward to also check ES relations for higher-point integrals/amplitudes. The two-loop double-pentagon integrals  $\mathcal{I}_{\text{dp}}(i,j,k,l)$ are all we need for MHV amplitudes (after regularizing divergences) as well as those finite components of NMHV amplitudes. Again the ``rational" part of $\mathcal{I}_{\text{dp}}(i,j,k,l)$ contains certain violation of ES relations at second and third entries, which require contributions from ``irrational" part. What we find is that for each of the $16$ four-mass-box square roots (depending on four dual points $x_a, x_b, x_c, x_d$)~\footnote{For ${\cal I}_{\rm dp}(i,j,k,l)$, generically we have $16$ choices for the dual points $(x_a, x_b, x_c, x_d)$ with $a=i, i{+}1$, $b=j,j{+}1$, $c=k,k{+}1$, $d=l, l{+}1$.}, the double discontinuity around $x_{a,c}^2=0$ and $x_{b,d}^2=0$ is again cancelled nicely. Thus $\mathcal{I}_{\text{dp}}(i,j,k,l)$  satisfies ES relations in exactly the same way.

An immediate consequence of this observation is that for $n\geq 8$,  all $\eta_i\eta_j\eta_k\eta_l$ components of two-loop NMHV amplitudes with non-adjacent $i<j<k<l$ also satisfy ES relations, since any such component is given by $\mathcal{I}_{\text{dp}}(i,j,k,l)-\mathcal{I}_{\text{dp}}(j,k,l,i)$. In fact, these finite components are absent of algebraic letters, so ES relations for these components is nothing but the absence of adjacent unfrozen letters $\langle a{-}1\,a\,b{-}1\,b\rangle$ and $\langle c{-}1\,c\,d{-}1\,d\rangle$ for $(a,b)\sim (c,d)$. This is a direct confirmation of ES relations to all multiplicities for finite amplitudes without any subtraction.

\paragraph{Three-loop MHV amplitudes and cluster adjacency} Given that finite double-pentagon integrals satisfy ES relations, so do two-loop MHV amplitudes if one can subtract IR divergences in a suitable way. In~\cite{Golden:2019kks}, cluster adjacency conditions have been checked for amplitudes with BDS-like subtraction or minimal subtraction when $n$ is a multiple of four, which imply ES relations when applied to letters of the form $\langle i{-}1 i j{-}1 j\rangle$. Here we move to the three-loop $n=8$ MHV amplitude, whose symbol has only been computed very recently~\cite{Li2021} using ${\bar Q}$ equations and two-loop $n=9$ NMHV amplitudes. This was a very non-trivial calculation: the raw data consists of roughly $10^9$ terms and further simplification is still underway. The alphabet consists of $204$ rational letters and $2 \times 9$ algebraic ones, and quite remarkably it still has the nice feature that algebraic letters only appear at second and third entries, just as two-loop NMHV amplitudes and ladder integrals.

First we need to convert this result to the one with minimal subtraction:
\[
\mathcal S(E_8^{(3)})=\mathcal S\left(R_8^{(3)}+\frac{F_8^3}{6} + R_8^{(2)}F_8\right),
\]
where $R_8^{(L)}$ are $L$-loop BDS-normalized amplitudes and $F_8$ is defined in the appendix \ref{A}. Despite the huge size of the symbol, we can check ES relations exactly as before, and we find the same phenomenon as in $I_{\text{dp}}^{(L)}$ and $\mathcal{I}_{\text{dp}}$. The rational part of $\mathcal S(E_8^{(3)})$ naively do not satisfy the extended Steinmann relations, which are broken at overlapping channels for each of the two four-mass boxes. We have only found such violation in the second and third entries, which is nicely in accordance with the position of algebraic letters. Quite remarkably, once we put these two parts together, the whole $\mathcal S(E_8^{(3)})$ satisfies the extended Steinmann relations as in previous examples!

Note that for $n=6,7$ cases, it is remarkable that ES relations turn out to be equivalent to cluster adjacency conditions, which states that any two adjacent letters as cluster variables of the finite cluster algebra $G(4,n)$ must belong to the same cluster. 
Given that all integrals and amplitudes we have checked satisfy ES relations, it is natural to ask if they also satisfy cluster adjacency conditions. Note that for $n\geq 8$, the relation between cluster adjacency and ES relations is unclear, except that for rational letters the latter follow from a special case of the former. It has been proven \cite{weaksep} that two unfrozen Mandelstam-type $\mathcal{A}$-coordinates $\langle i{-}1 i j{-}1 j \rangle$
and $\langle k{-}1 k l{-}1 l\rangle$ are cluster adjacent if and only if $(i, j)\sim (k, l)$ in the $n$-gon.

However, starting from $n=8$, there are two main obstructions for studying cluster adjacency. First, it is known that not all letters are cluster variables of $G(4,n)$: algebraic letters or even more complicated functions can appear. Second, $G(4,n)$ for $n\geq 8$ is no longer a finite cluster algebra, which has infinite cluster variables, which makes it difficult to check whether a rational letter is a cluster variable and whether two cluster variables belong to a same cluster.

In \cite{Golden:2019kks}, Sklyanin bracket \cite{Sklyanin:1983ig} was introduced to resolve the second problem. Sklyanin bracket is a anti-symmetric Poisson bracket of two functions on $G(4,n)$. Suppose we parametrize $G(4,n)$ by 
\[
Z=\begin{pmatrix}
1 & 0 & 0 & 0 & y_{5}^{1} & \ldots & y_{n}^{1} \\
0 & 1 & 0 & 0 & y_{5}^{2} & \ldots & y_{n}^{2} \\
0 & 0 & 1 & 0 & y_{5}^{3} & \ldots & y_{n}^{3} \\
0 & 0 & 0 & 1 & y_{5}^{4} & \ldots & y_{n}^{4}
\end{pmatrix},
\]
then the Sklyanin bracket of two coordinates are given by 
\begin{equation}
\left\{y_{a}^{I}, y_{b}^{J}\right\}=\frac{1}{2}(\operatorname{sign}(J-I)-\operatorname{sign}(b-a)) y_{a}^{J} y_{b}^{I},
\end{equation}
and it is defined for any two functions by
\begin{equation}
\{f(y), g(y)\}=\sum_{a, b=5}^{n} \sum_{I, J=1}^{4} \frac{\partial f}{\partial y_{a}^{I}} \frac{\partial g}{\partial y_{b}^{J}}\left\{y_{a}^{I}, y_{b}^{J}\right\}.
\end{equation}
The main conjecture from \cite{Golden:2019kks} is: \textit{two $\mathcal{A}$-coordinates $a_1$, $a_2$ exist in the same cluster (thus cluster adjacent) if and only if their Sklyanin bracket $\{\log a_1, \log a_2\}$ is an integer or a half integer, i.e. $\{\log a_1, \log a_2\}\in \frac{1}{2}\mathbb Z$}. 

For the first problem, since algebraic letters are no longer cluster $\mathcal{A}$-coordinates or $\mathcal{X}$-coordinates, we do not know a notion of cluster adjacency concerning them. Therefore, this check can only be performed for all pairs of rational letters. Before we apply it to more non-trivial cases, we first use this conjecture to check cluster adjacency conditions of $I_{\text{pb}}^{(L)}$, whose DCI alphabet forms a $D_3$ cluster algebra. By expanding the alphabet in $\mathcal{A}$-coordinates, we find that it becomes the union of
\[
\bigl\{\langle 812(67)\cap(345)\rangle,\langle 6781\rangle,\langle 1467\rangle,\langle 4567\rangle,\langle 8124\rangle,\langle 1234\rangle,\langle 1345\rangle\bigr\}
\]
and
\begin{align*}
\bigl\{&\langle 1 (28)(34)(67)\rangle,\, 
\langle 1 (28)(45)(67)\rangle,\, 
\langle 4 (18)(35)(67)\rangle,\, 
\langle 4 (12)(35)(67)\rangle,\\
&
\langle 1267\rangle,
\langle 3467\rangle,
\langle 1348\rangle,
\langle 1458\rangle,
\langle 1245\rangle\bigr\}.
\end{align*}
By computing the Sklyanin brackets for all pairs of letters, we find that only the following 15 pairs, which are exclusively contained in the second list, are not in $\frac{1}{2}\mathbb Z$, which means that they cannot appear next to each other in the symbol of $I_{\text{pb}}^{(L)}$:
\begin{align*}
&(\langle 1 (28)(34)(67)\rangle,\langle 4 (81)(35)(67)\rangle),\,\,
(\langle 1 (28)(34)(67)\rangle,\langle 4 (12)(35)(67)\rangle),\\
&(\langle 1 (28)(34)(67)\rangle,\langle 1458\rangle),\,\,
(\langle 1 (28)(34)(67)\rangle,\langle 1245\rangle),\\
&(\langle 1 (28)(45)(67)\rangle,\langle 4 (81)(35)(67)\rangle),\,\,
(\langle 1 (28)(45)(67)\rangle,\langle 4 (12)(35)(67)\rangle),\\
&(\langle 1 (28)(45)(67)\rangle,\langle 3467\rangle),\,\,
(\langle 4 (18)(35)(67)\rangle,\langle 1267\rangle),\,\,
(\langle 4 (12)(35)(67)\rangle,\langle 1348\rangle),\\
&(\langle 4 (12)(35)(67)\rangle,\langle 1458\rangle),\,\,
(\langle 1267\rangle,\langle 1348\rangle),\,\,
(\langle 1267\rangle,\langle 1458\rangle),\,\,
(\langle 3467\rangle,\langle 1458\rangle),\\
&(\langle 3467\rangle,\langle 1245\rangle),\,\,
(\langle 1348\rangle,\langle 1245\rangle)\,.
\end{align*}
We have checked this is indeed the case up to $L=4$. Similar computation shows that the alphabet of $I_{dp}^{L}(1,4,5,7)$ contains $7+16$ $\mathcal{A}$-coordinates which are
\[
\bigl\{\langle 1234\rangle, \langle 1247\rangle, \langle 1267\rangle, \langle 1345\rangle, 
 \langle 1567\rangle, \langle 3456\rangle, \langle 4567\rangle\bigr\}
\]
and
\begin{align*}
\bigl\{&\langle 1(27)(34)(56)\rangle, 
 \langle 4(12)(35)(67)\rangle, 
 \langle 5(12)(34)(67)\rangle, 
 \langle 6(12)(34)(57)\rangle, 
 \langle 7(12)(34)(56)\rangle, \\
 &\langle 1245\rangle, \langle 1246\rangle, 
 \langle 1256\rangle, \langle 1257\rangle, \langle 1346\rangle, \langle 1347\rangle, 
 \langle 1456\rangle, \langle 1457\rangle, \langle 1467\rangle, \langle 3457\rangle, 
 \langle 3467\rangle\bigr\}
\end{align*}
Very nicely, we find again that $54$ pairs of letters have Sklyanin brackets that are not in $\frac{1}{2}\mathbb Z$, which only concern the $16$ letters in the second list. These are the pairs that cannot appear next to each other in the symbol $\mathcal{S}(I_{dp}^{L}(1,4,5,7))$, which we have also confirmed explicitly up to $L=3$.

Note that a prior these cluster adjacency conditions based on (conjectural) Sklyanin bracket are different from the notion of adjacency for any pair of $F$ polynomials in a cluster-algebra alphabet as studied in~\cite{He:2021non}. However, we find that for the $D_4$ case, the latter is a consequence of the former! In other words, any two $F$ polynomials cannot appear to each other if and only if they appear in a forbidden pair of $A$ coordinates (with Sklyanin bracket not in $\frac12 \mathbb{Z}$). 

Next we apply this conjecture to check cluster adjacency conditions of integrals and amplitudes with algebraic letters. For $I_{\text{dp}}^{(L)}(1,4,5,8)$, there is a pair $\langle 1256\rangle$ and $\langle 3478\rangle$ that appear in the second and third entries, which naively violate cluster adjacency. However, this must be an artifact of focusing on rational letters only, and the naive violation should be cancelled by contributions from algebraic letters (though we have not defined adjacency for the latter) similar to the case for ES relations. The non-trivial observation is that all other pairs of $\mathcal{A}$-coordinates in consecutive entries of the symbol indeed have brackets in $\frac{1}{2}\mathbb Z$, thus cluster adjacency is respected for them according to the conjecture. 

We can proceed and check cluster adjacency conditions for finite integrals and component amplitudes, as well as three-loop $n=8$ MHV amplitudes with minimal subtraction. Remarkably, except for the letters corresponding to overlapping channels in four-mass boxes at entry $(2,3)$, all other pairs of $\mathcal{A}$-coordinates in consecutive entries have brackets in $\frac{1}{2}\mathbb Z$. Provided the conjecture~\cite{Golden:2019kks}, we have thus confirmed cluster adjacency conditions for all cases we considered, except for the obvious violation which should be saved by algebraic letters. Naively these cluster adjacency conditions seem much stronger than ES relations but this can change if we impose them for bootstrap; we leave a systematic study of ES relations/cluster adjacency for octagon bootstrap (including algebraic letters), to future works. 

\section{Last-two-entries for MHV amplitudes from ${\bar Q}$ equations}

Finally, let us turn to the discussion of the last-two-entries for $L$ loops $n$-points MHV amplitude. Unlike extended Steinmann relations which are expected to hold for any (finite) integrals and amplitudes, a prior we do not expect such universal constraints on the last entry (let alone last two entries). However, for MHV amplitudes with BDS normalization, it has been known since~\cite{CaronHuot:2011ky} that dual supersymmetry $\bar{Q}$ restrict the last entry to be of the form $\langle \bar{i} j\rangle$ ($n(n{-}5)$ unfrozen variables in total), and by studying the RHS of ${\bar Q}$ equations for MHV amplitude a bit more carefully, we can fix the last two entries as well. 

We will first quickly review basic ingredients of $\bar Q$ formalism  \cite{CaronHuot:2011kk}, which offers an efficient method to calculate amplitudes of higher loops from the lower-loop ones. Equipped with the enumeration of all possible last entries (times R invariants) of NMHV~\cite{He:2020vob}, it becomes possible to use ${\bar Q}$ equations to determine all possible last-two-entries for MHV amplitudes. To do so, it is important to use the symbol integration algorithm presented in \cite{CaronHuot:2011kk}, which computes the symbol of $d\log$ integral purely from the symbol of the integrand. This allows us to convert those ``last entry $\times$ R invariants", into last-two-entries for MHV amplitude. This can be done for any $n$, but we will mostly focus on $n=6,7,8,9$, which suffices to illustrate our method.

 As proposed in \cite{CaronHuot:2011kk}, $\bar Q$ anomaly of the $L$-loop {\it BDS-normalized} $S$-matrix $R_{n,k}^{(L)}$ is related to those of higher $n$ and $k$ but lower $L$, according to the perturbative expansion of the $\bar Q$ equation. Since kernel of $\bar Q$ is trivial for $k=0,1$, after the anomaly $\bar Q_a^A R^{(L)}_{n,k}$ is determined, we can directly replace $\bar Q_a^A$ by ${\rm d}$ to get the total derivative ${\rm d}R^{(L)}_{n,k}$, hence the symbol $\mathcal{S}(R^{(L)}_{n,k})$. This makes $\bar Q$ equation a powerful tool for computing NMHV (MHV) amplitudes from N$^2$MHV  (NMHV) amplitudes, and even more so for determining (all-loop) last entries. The ${\bar Q}$ anomaly of MHV amplitudes,can be obtained from $(n{+}1)$-point NMHV amplitudes:
\begin{equation}
    R_{n{+}1,1}^{(L)}=\sum_{1\leq j<k<l<m\leq n}[j\ k\ l\ m\ n{+}1]F_{j,k,l,m}^{(L)}(Z)
\end{equation}
where $F_{j,k,l,m}^{(L)}(Z)$ are weight $2 L$ multi-polylog functions. According to the $\bar Q$ equations, after taking the collinear limit ${Z}_{n{+}1}={Z}_n-\epsilon{Z}_{n{-}1}+C\epsilon\tau{Z}_1+C^\prime\epsilon^2{Z}_2$ with $C=\frac{\langle n{-}1n23\rangle}{\langle n123\rangle}$ and $C^\prime=\frac{\langle n{-}2n{-}1n1\rangle}{\langle n{-}2n{-}121\rangle}$ in $R_{n{+}1,1}^{(L)}$, formally we are performing integrals as
\begin{equation}\label{toqlog1}
     \text{Res}_{\epsilon=0}\int {\rm d}^{2|3}\mathcal{Z}_{n{+}1}[j\ k\ l\ m\  n{+}1]F(\epsilon,\tau)
\end{equation}
which allows us to obtain ${\bar Q}$ (thus total differential) of $n$-point MHV amplitudes. General $F(\epsilon,\tau)$ will lead to divergences when taking the residue at $\epsilon=0$. However, such divergences cancel in the combination of RHS of ${\bar  Q}$ equations, and all we need are finite part of $F$'s in the limit $\epsilon\to0$. (for general $L$ we do not know or need explicit form of these $F$'s). 
The upshot is~\cite{CaronHuot:2011kk}:
\begin{align}\label{toqlog2}
    \text{Res}_{\epsilon=0}&\int^{\tau=\infty}_{\tau=0} {\rm d}^{2|3}\mathcal{Z}_{n{+}1}[i\,j\,k\,n\,n{+}1]F(\epsilon,\tau)\nonumber\\
    &=\int^{\infty}_0\biggl({\rm d}\log\frac{\langle Xij\rangle}{\langle Xjk\rangle}\bar Q\log\frac{\langle\bar n j\rangle}{\langle\bar ni\rangle}+{\rm d}\log\frac{\langle Xjk\rangle}{\langle Xik\rangle}\bar Q\log\frac{\langle\bar n k\rangle}{\langle\bar ni\rangle}\biggr)F(\epsilon,\tau)\biggm|_{\epsilon\to0\ \text{finite part}}
\end{align}
for $i,j,k\neq1,n{-}1$, and similarly
\begin{align}
[i\ j\ n{-}1\ n\ n{+1}]&\to\int{\rm d}\log\frac{\langle Xij\rangle}{\langle X n{-}2 n{-}1\rangle}\bar Q\log \frac{\langle\bar n j\rangle}{\langle\bar ni\rangle}\nonumber\\
[1\ i\ j\ n\ n{+1}]&\to\int{\rm d}\log\frac{\langle Xij\rangle}{\langle X 12\rangle}\bar Q\log \frac{\langle\bar n j\rangle}{\langle\bar ni\rangle}\nonumber\\
[1\ i\ n{-}1\ n\ n{+1}]&\to\int{\rm d}\log\frac{\langle X n{-}2 n{-}1\rangle}{\langle X12\rangle}\bar Q\log \frac{\langle\bar n j\rangle}{\langle\bar ni\rangle}
\end{align}
with $X=n\wedge B$ and $\mathcal{Z}_B=\mathcal{Z}_{n{-}1}-C\tau\mathcal{Z}_1$. The integrals vanish for other R-invariants.

Practically, these integrations are performed at the symbol level following the algorithm presented in the appendix of \cite{CaronHuot:2011kk}. Last entries of the integrand together with the ${\rm d}\log$ form fully determine the last-two-entries for the result. Nicely, all last entries for NMHV amplitudes have been given in \cite{He:2020vob}, which we list in Appendix \ref{lastentryNMHV} for completeness. This is our starting point for computing all possible last-two-entries for MHV amplitudes.

Now we present some details of our calculation for possible last-two-entries of $n$-point MHV amplitudes. Following \eqref{toqlog1} and \eqref{toqlog2}, only terms  proportional to $[i\,j\,k\,n\,n{+}1]$ ($1\leq i<j<k\leq n{-}1$) contribute non-trivial results after performing the Grassmannian integration $({\rm d}^3\chi)^A$ and taking the residue at $\epsilon=0$. Hence we firstly list all the possible pairs $[i\,j\,k\,n\,n{+}1]F(\epsilon,\tau)\otimes w(\epsilon,\tau)$, where $[i\,j\,k\,n\,n{+}1]w(\epsilon,\tau)$ are NMHV last entries and $F(\epsilon,\tau)$ are arbitrary unknown entries, both after applying the collinear limit. Secondly, we perform the Grassmannian integration over the Yangian part of these pairs following \eqref{toqlog1} and \eqref{toqlog2}. To take the finite part at $\epsilon=0$, we can simply set overall $\epsilon^k$ factors of $w(\epsilon,\tau)$ to $1$ and other $\epsilon$ to $0$. Finally, we integrate $\tau$ on $\mathbb{R}_{\geq0}$, obtaining a list of symbols associated with $\bar Q\log\frac{\langle\bar ni\rangle}{\langle\bar nj\rangle}$. We record last-two-entries of this list, which gives an all-loop prediction for MHV amplitudes.

\paragraph{$\bf{n=6}$}
We first take $n=6$ MHV as an illustration. There are $10$ R-invariants of $7$-point NMHV which give non-trivial contributions. For instance, $[13567]$ is associated with $\bar Q\log\langle\bar ni\rangle$, which are in the first class of NMHV last entries, and a non-trivial one $\bar Q\log\frac{\langle3(42)(56)(71)\rangle}{\langle2345\rangle\langle1367\rangle}$ is in the second class. For the latter one, we take $[13567]\to{\rm d}\log\frac{\langle X45\rangle}{\langle X12\rangle}\bar Q\log\frac{\langle\bar 62\rangle}{\langle\bar63\rangle}$ and parametrize $\mathcal{Z}_7$ in collinear limit. $\frac{\langle3(42)(56)(71)\rangle}{\langle2345\rangle\langle1367\rangle}$ then factorizes out an $1/\epsilon$ factor which is set to $1$ while other $\epsilon\to0$. Finally, after performing the $\tau$-integration we get eight possible pairs
\[
\{\langle1236\rangle,\langle2346\rangle,\langle1235\rangle,\langle2345\rangle\}\otimes \bar Q\log\frac{\langle\bar62\rangle}{\langle\bar63\rangle}\]
as last-two-entries for $n=6$ MHV amplitudes from this contribution.

Following the same logic and collect all possible contributions, we find that $\bar Q$ equations in fact leave no constraints for possible last-two-entries of $n=6$, {\it i.e.} for each $\langle\bar i j\rangle$ as last entries, all $15$ $\mathcal{A}$-coordinates are allowed to appear before it. It will not be that case for higher $n$ however, and we will see that the $\bar Q$ equation impose stronger constraints as $n$ increases.

\paragraph{$\bf{n=7}$}
For $7$-point MHV amplitudes, there are two independent unfrozen last entries up to cyclicity, which are chosen as $\langle1367\rangle$ and $\langle1467\rangle$. Naive computation shows that there are both $35$ possible $\mathcal{A}$-coordinates allowed to appear before them. However, since $\langle1367\rangle$ and $\langle1467\rangle$ are related by the dihedral symmetry $\{3\leftrightarrow4,2\leftrightarrow5,1\leftrightarrow6\}$, letters before them should also preserve such a symmetry. Thus we need to delete all those $\mathcal{A}$-coordinates before $\langle1367\rangle$ that are not dihedral images of those before $\langle1467\rangle$ and {\it vice versa}, ruling out $10$ letters in each list. 

Moreover, for $7$-point amplitudes we have the canonical choice of the $42$ symbol letters in the DCI form
\begin{align}
a_{11}=\frac{\langle1234\rangle\langle1567\rangle\langle2367\rangle}{\langle1237\rangle\langle1267\rangle\langle3456\rangle},\ \ a_{21}=\frac{\langle1234\rangle\langle2567\rangle}{\langle1267\rangle\langle2345\rangle}, \ \ a_{31}=\frac{\langle1567\rangle\langle2347\rangle}{\langle1237\rangle\langle4567\rangle},\nonumber\\
a_{41}=\frac{\langle2457\rangle\langle3456\rangle}{\langle2345\rangle\langle4567\rangle},\ \ 
a_{51}=\frac{\langle1(23)(45)(67)\rangle}{\langle1234\rangle\langle1567\rangle},\ \ 
a_{61}=\frac{\langle1(34)(56)(72)\rangle}{\langle1234\rangle\langle1567\rangle},
\end{align}
together with all $a_{ij}$ from cyclically relabeling $a_{i1}$ by $j{-}1$ times.  Each $a_{ij}$ contains only one unfrozen $\mathcal{A}$-coordinates, such that every pair of unfrozen letters is in one-to-one correspondence with a pair like $a_{ij}\otimes a_{kl}$. Hence we can recover DCI by simply replacing each unfrozen $\mathcal{A}$-coordinates by its corresponding $a_{ij}$. Note that the $14$ unfrozen last entries $\langle\bar ij\rangle$ correspond to $\{a_{2i},a_{3i}\}$ for $i=1\cdots7$.

In this way, we find exactly $25$ allowed letters before $a_{22}=\frac{\langle1367\rangle\langle2345\rangle}{\langle1237\rangle\langle3456\rangle}$, which are
\[\{a_{11},a_{12},a_{15},a_{16},a_{17},a_{2i},a_{3i},a_{42},a_{44},a_{46},a_{47},a_{56},a_{57}\}\]
with $i=1\cdots7$. Letters before other last entries can be obtained by cyclic rotation and dihedral symmetries. In total there are $350$ DCI pairs and we have checked that our prediction matches the explicit results of $7$-point MHV amplitudes up to $L=3$, for both the BDS-subtracted amplitude \cite{Drummond:2014ffa}  and the one after BDS-like subtraction \cite{Dixon:2016nkn}. Note that our list does not preserve extended Steinmann relations since we are using standard BDS-subtraction. However,at $n=6,7$ the BDS-like subtraction \cite{Alday:2009dv} only reduces the possible pairs for last-two-entries comparing to BDS-subtraction. Therefore our prediction still holds for amplitudes with BDS-like subtraction.

Furthermore, we reorganize these $350$ weight-$2$ symbols by applying the integrability condition $\sum_{i,j,k,l} {\rm d}\log(a_{ij}){\rm d}\log(a_{kl})=0$, and conclude that there are $139$ independent functions. $(13\times14)/2+14=105$ of them are of the form $\log(x)\log(y)$, where $x,y\in\{a_{2i},a_{3i}\}_{i=1\cdots7}$. And the remaining $34$ functions live in $5$ cyclic orbits, whose initial seeds can be chosen as
\begin{equation*}
\biggl\{\Li_2\biggl(\frac{a_{31}}{a_{25}a_{26}}\biggr),\Li_2\biggl(\frac{a_{21}}{a_{33}a_{34}}\biggr),\Li_2(a_{21}a_{35}),\Li_2(a_{21}a_{32}),\Phi \biggr\}
\end{equation*}
where the symbol of the last one reads~\footnote{The weight-$2$ function $\Phi$ can be easily obtained but we suppress the expression since it is much longer than the symbol.}:
\[\mathcal{S}(\Phi)=-(a_{31}a_{33})\otimes a_{25}+a_{36}\otimes (a_{21}a_{23})+a_{41}\otimes\frac{a_{31}a_{36}}{a_{21}a_{23}a_{26}}-a_{43}\otimes\frac{a_{31}a_{33}}{a_{21}a_{23}a_{25}}\]
We remark that the length of the last orbit is only $6$ (thus there are $7+7+7+7+6=34$ functions in total), since after cyclic rotation $6$ times the image turns out to be a linear combination of the first $6$ functions in this orbit. This gives all integrable weight-$2$ symbols for the last two entries.

\paragraph{$\bf{n\geq8}$}
It is straightforward to extend the computation to any $n$, but since the simplification of the $\mathcal{A}$-coordinates becomes more and more complicated, the list becomes more and more intricate as $n$ grows. We will not attempt to find such lists for  all multiplicities, but content ourselves with last-two-entries for $n=8,9$, which are generic enough. Note that since we do not have a confirmed alphabet beyond $n=6,7$, it is not clear how they can be organized into DCI letters or how to apply integrability conditions without any ambiguities. 

For $n=8$, we have two independent unfrozen last entries $\langle1378\rangle$ and $\langle1478\rangle$ up to cyclic shift and dihedral symmetry $\{3\leftrightarrow4,2\leftrightarrow5,1\leftrightarrow6,7\leftrightarrow8\}$; explicit computation shows that there are $45$ unfrozen $\mathcal{A}$-coordinates before $\langle1378\rangle$ and $46$ before $\langle1478\rangle$, which are recorded in the appendix \ref{C}. Note that this represents a significant reduction of the weight-$2$ symbols in the last two entries, as naively there can be order $200$ letters for any last entry. We have checked that our prediction holds for $8$-point BDS-subtracted amplitudes up to $L=3$~\cite{Li2021}.

For $n=9$ there are also $2$ independent unfrozen last entries, {\it e.g.} $\langle1389\rangle$, $\langle1489\rangle$ up to cyclic shift and dihedral symmetries. There are $70$ allowed unfrozen letters before $\langle1389\rangle$ and $71$ before $\langle1489\rangle$. As discussed in \cite{Henke:2021avn,Ren:2021ztg} using the tropical Grassmannian, naively there can be thousands of rational letters before each last entry. Our computation, however, drastically reduces this number, showing that the possible pairs are in fact very limited. Thus we have seen a significant reduction of function space for $n>6$ from these last-two-entry conditions, and the fraction of reduction grows rapidly with $n$.
\begin{table}[t]
\hspace{-2.35ex}\begin{tabular}{|l|c|c|c|c|c|c|}
\hline
$n$ & $n=6$                           & $n=7$                           & \multicolumn{2}{c|}{$n=8$} & \multicolumn{2}{c|}{$n=9$} \\ \hline
independent (unfrozen) $\langle\bar ni\rangle$ & $\langle1356\rangle$ & $\langle1367\rangle$ & $\langle1378\rangle$        & $\langle1478\rangle$       & $\langle1389\rangle$        & $\langle1489\rangle$       \\ \hline
num. of allowed (unfrozen) letters & 9                             & 25                            & 45          & 46         & 70          & 71         \\ \hline
\end{tabular}
\caption{Numbers of last-two-entries for $n\leq9$ MHV amplitudes}\label{table1}
\end{table}

\section{Conclusion and Discussions}
In this paper, we have provided further evidence that extended-Steinmann relations hold for individual (finite) Feynman integrals as well as scattering amplitudes beyond $n=6,7$, including various ladder integrals up to four loops, two-loop double-pentagons and NMHV finite components to all multiplicities, as well as three-loop $n=8$ MHV amplitudes with minimal subtraction. In all cases with algebraic letters, the contribution from algebraic letters is needed for ES relations to hold, and we have also checked cluster adjacency for rational letters in terms of Sklyanin bracket. For the first two entries, we have also listed all $n(n{-}5)(n^2-n+5)/8$ dilogarithm functions which satisfy Steinmann relations. Finally, for MHV amplitudes we have computed all possible last two entries using ${\bar Q}$ equations, which provide further constraints on their function space, as summarized in the Table \ref{table1} above for up to $n=9$. 

Our result on ES relations provides new evidence for their validity for Feynman integrals and amplitudes beyond $n=6,7$: unlike MHV amplitudes where ES relations only hold with suitable normalization, individual Feynman integrals and finite components of non-MHV amplitudes do not require any subtraction thus ES relations hold in a ``cleaner" way; on the other hand, it is already non-trivial that ES relations hold for two-loop MHV amplitudes with minimal subtraction~\cite{Golden:2019kks}, and we find it remarkable that they hold for three-loop $n=8$ amplitudes in the same normalization. Our result has further supported the claim that ES relations is a general property for any QFT (at least in the planar limit in the way we have formulated it). It would be highly desirable to understand these relations better, especially for divergent integrals/amplitudes where subtraction or regularization is needed. It would be fascinating to see if one could identify some underlying principle responsible for these remarkable relations. 

Another mysterious property of some Feynman integrals and scattering amplitudes (in and beyond planar ${\cal N}=4$ SYM) is that their symbol alphabet seems to fit into a cluster algebra (or truncated ones~\cite{He:2021non} with algebraic letters). In the presence of such cluster algebras, ES relations become closely related to cluster adjacency conditions, and we have established cluster adjacency conditions using $A$-coordinates in all these integrals and amplitudes, modulo subtlety with algebraic letters. It would be highly desirable to formulate similar (truncated) cluster adjacency conditions including the algebraic letters, and to understand connections with ES relations {\it e.g.} for $n=8$ case. These ES relations/cluster adjacency can be used as extremely powerful constraints for bootstrap, at least for $n=8$ case which seems within reach. The starting point is the space of weight-$2$ functions we have constructed, and it would be interesting to continue building ES/adjacency-satisfying space at higher weights, which can be used for bootstrapping multi-loop amplitudes/integrals for $n=8$.

Last but not least, although we have focused on studying ES relations for multiple polylogs (MPL) at the symbol level, as constraints on iterative discontinuities they apply to more general functions such as elliptic MPL. For example, it is clear that the fully massive double-box integral (finite component of two-loop $n=10$ N${}^3$MHV amplitudes)~\cite{Kristensson:2021ani} satisfies Steinmann relations, and we leave the study of ES relations for integrals/amplitudes beyond MPL to future works. 

\section*{Acknowledgement}
It is a pleasure to thank Yichao Tang and Chi Zhang for collaborations on related projects. This research is supported in part by National Natural Science Foundation of China under Grant No. 11935013,11947301, 12047502,12047503.

\appendix

\section{Minimally-normalized Amplitude}\label{A}

The minimally-normalized (MHV) amplitude $E_n$ is defined by
$$
E_n=R_n \exp\left(\frac{\Gamma_{\text{cusp}}}{4} F_n\right),
$$
where $R_n$ is the BDS-normalized amplitude, then 
$$
\begin{aligned}
E_n&=(1 + R_n^{(2)} g^4 + R_n^{(3)} g^6 + \cdots)\exp\left((g^2-2\zeta(2)g^4+22\zeta(4)g^6+\cdots) F_n\right)\\
&=1+F_n g^2+g^4 \left(R_n^{(2)}+\frac{F_n^2}{2}-2 \zeta(2) F_n\right)\\
&\quad +g^6\biggl(R_n^{(3)}+\frac{F_n^3}{6}-2 \zeta(2) F_n^2 + (22 \zeta(4)+R_n^{(2)})F_n\biggr)+\cdots
\end{aligned}
$$
The function $F_n$ is defined by
\begin{equation*}
F_n=\sum_{i=1}^{n}\biggl[f(i)+\sum_{j=2}^{\bigl\lfloor\frac{n-3}{2}\bigr\rfloor} h(i, j)\biggr]+
\sum_{i=1}^n
\begin{cases}
g(i, n/2) & \text{$n$ even};\\
0 & \text{$n$ odd},
\end{cases}
\end{equation*}
where
$$
\begin{aligned}
f(i)&=\frac{3}{2} \zeta(2)+\frac{1}{2} \log ^{2}\left(\frac{x_{i-1,i+2}^2}{x_{i,i+2}^2}\right)-\log \left(\frac{x_{i,i+2}^2}{x_{i,i+3}^2}\right) \log \left(\frac{x_{i+1,i+3}^2}{x_{i,i+3}^2}\right),\\
g\left(i, k\right)&=\frac{1}{2}  \operatorname{Li}_{2}\left(1-\frac{x^2_{i-1, i+k-1} x^2_{i, i+k}}{x^2_{i-1, i+k} x^2_{i, i+k-1}}\right) +\frac{1}{2} \log ^{2}\left(\frac{x^2_{i, i+k-1}}{x^2_{i, i+k}}\right)-\frac{1}{4} \log ^{2}\left(\frac{x^2_{i-1, i+k}}{x^2_{i, i+k-1}}\right) \\
&\quad-\frac{1}{2} \log \left(\frac{x^2_{i-1, i+k-1} x^2_{i, i+k}}{x^2_{i-1, i+k} x^2_{i, i+k-1}}\right) \log \left(\frac{x^2_{i-1, i+k}}{x^2_{i, i+k-1}}\right),\\
h(i, j)&=\operatorname{Li}_{2}\left(1-\frac{x^2_{i-1, i+j} x^2_{i, i+j+1}}{x^2_{i-1, i+j+1} x^2_{i, i+j}}\right)+\log \left(\frac{x^2_{i, i+j+1}}{x^2_{i-1, i+j+1}}\right) \log \left(\frac{x^2_{i, i+j+1}}{x^2_{i, i+j}}\right)
\end{aligned}
$$
where $x_{i,j}^2$ breaks dual conformal symmetry.
For our purpose, we write the symbol of $E_n$ from that of $R_n$ and $F_n$ as
\begin{equation*}
\mathcal S(E_n^{(1)})=\mathcal S(F_n),\quad 
    \mathcal S(E_n^{(2)})=\mathcal S\left(R_n^{(2)}+\frac{F_n^2}{2}\right),\quad 
    \mathcal S(E_n^{(3)})=\mathcal S\left(R_n^{(3)}+\frac{F_n^3}{6} + R_n^{(2)}F_n\right).
\end{equation*}

\section{Last-entries for all-multiplicity NMHV amplitudes}\label{lastentryNMHV}
Let us summarize all possible last entries (dressed with Yangian invariants) for $n$-point NMHV amplitudes~\cite{He:2020vob}, where we have three classes (note the second term on the right hand side of $\bar Q$ equations also belongs to the first class):
\begin{itemize}
    \item [1.] $(n{-}4)\binom{n-1}{4}-\binom{n-3}{2}$ last entries: \begin{equation}[abcde]\bar Q\log\frac{\langle\bar ni\rangle}{\langle\bar ni\rangle},\end{equation}
    \item [2.] $3\binom{n-3}{4}\,$ last entries: 
    \begin{align}
    &[1\,i_{1}\,i_{2}\,i_{3}\,i_{4}]\,\bar{Q}\log\frac{\langle1(n{-}1\,n)(i_{1}\,i_{2})(i_{3}\,i_{4})\rangle}{\langle\bar{n}i_{1}\rangle\langle1i_{2}i_{3}i_{4}\rangle} \:,  \nonumber \\
    &[i_{1}\,i_{2}\,i_{3}\,i_{4}\,n{-}1]\,\bar{Q}\log\frac{\langle n{-}1(n\,1)(i_{1}\,i_{2})(i_{3}\,i_{4})\rangle}{\langle\bar{n}i_{1}\rangle\langle n{-}1\,i_{2}i_{3}i_{4}\rangle} \label{lastentry2} \\
    &[i_{1}\,i_{2}\,i_{3}\,i_{4}\,n]\,\bar{Q}\log\frac{\langle n(1\,n{-1})(i_{1}\,i_{2})(i_{3}\,i_{4})\rangle}{\langle\bar{n}i_{1}\rangle\langle n\,i_{2}i_{3}i_{4}\rangle}  \nonumber
\end{align}
where $\,1<i_{1}<i_{2}<i_{3}<i_{4}<n{-}1$, and the abbreviation $\langle a(bc)(de)(fg)\rangle:=\langle abde\rangle\langle acfg\rangle-\langle acde\rangle\langle abfg\rangle$.
    \item [3.] $\,2\binom{n-3}{5}\,$ last entries: 
    \begin{equation} \label{lastentry3}
    [i_{1}\,i_{2}\,i_{3}\,i_{4}\,i_{5}]\,\bar{Q}\log\frac{\langle\bar{n}(i_{1}i_{2})\cap(i_{3}i_{4}i_{5})\rangle}{\langle\bar{n}i_{1}\rangle\langle i_{2}i_{3}i_{4}i_{5}\rangle} \:, \:\: [i_{1}\,i_{2}\,i_{3}\,i_{4}\,i_{5}]\,\bar{Q}\log\frac{\langle\bar{n}(i_{1}i_{2}i_{3})\cap(i_{4}i_{5})\rangle}{\langle\bar{n}i_{1}\rangle\langle i_{2}i_{3}i_{4}i_{5}\rangle} \:,
\end{equation}
where $\,1<i_{1}<i_{2}<i_{3}<i_{4}<i_{5}<n{-}1$.
\end{itemize}
For computing $n$-point MHV amplitudes using ${\bar Q}$ equations, we take $n\to n{+}1$ and plug these on the RHS of ${\bar Q}$ equations for MHV.  
\section{Last-two-entries for $n=8$ MHV amplitudes}\label{C}
For $n=8$, there are two independent unfrozen last entries up to cyclicity and dihedral symmetries,$\langle1378\rangle$ and $\langle1478\rangle$. We list allowed unfrozen $\mathcal{A}$-coordinates 
\begin{align*}
    \{\langle 7 (18)(23)(56)\rangle, 
 \langle 7 (18)(34)(56)\rangle, 
 \langle 8 (12)(34)(67)\rangle, 
 \langle 8 (12)(35)(67)\rangle, 
 \langle 2 3 (178)\cap (456)\rangle,\\
 \langle 1235\rangle, \langle 1236\rangle,
 \langle 1237\rangle, \langle 1248\rangle, \langle 1258\rangle, \langle 1267\rangle, 
 \langle 1268\rangle, \langle 1345\rangle, \langle 1347\rangle, \langle 1348\rangle, \\
 \langle 1357\rangle, \langle 1358\rangle, \langle 1367\rangle, \langle 1368\rangle, 
 \langle 1378\rangle, \langle 1456\rangle, \langle 1478\rangle, \langle 1567\rangle, 
 \langle 1568\rangle, \langle 1578\rangle,\\
 \langle 2346\rangle, \langle 2347\rangle, 
 \langle 2348\rangle, \langle 2358\rangle, \langle 2368\rangle, \langle 2378\rangle, 
 \langle 2456\rangle, \langle 2567\rangle, \langle 2678\rangle,
 \langle 3457\rangle,\\ 
 \langle 3458\rangle, \langle 3468\rangle, \langle 3478\rangle, \langle 3567\rangle, 
 \langle 3568\rangle, \langle 3578\rangle, \langle 3678\rangle, \langle 4568\rangle, 
 \langle 4578\rangle, \langle 4678\rangle\}
\end{align*}
which can appear before $\langle1378\rangle$, and
\begin{align*}
    \{\langle 1 (23)(45)(78)\rangle, 
 \langle 1 (23)(46)(78)\rangle, 
 \langle 7 (18)(24)(56)\rangle, 
 \langle 7 (18)(34)(56)\rangle, 
 \langle 8 (12)(34)(67)\rangle,\\
 \langle 8 (12)(45)(67)\rangle, \langle 1235\rangle, \langle 1236\rangle, 
 \langle 1237\rangle, \langle 1247\rangle, \langle 1248\rangle, \langle 1258\rangle, 
 \langle 1267\rangle, \langle 1268\rangle,\\
 \langle 1345\rangle, \langle 1347\rangle, 
 \langle 1348\rangle, \langle 1378\rangle, \langle 1456\rangle, \langle 1457\rangle, 
 \langle 1458\rangle, \langle 1467\rangle, \langle 1468\rangle, \langle 1478\rangle,\\
 \langle 1567\rangle, \langle 1568\rangle, \langle 1578\rangle, \langle 2346\rangle, 
 \langle 2347\rangle, \langle 2348\rangle, \langle 2378\rangle, \langle 2456\rangle, 
 \langle 2458\rangle, \langle 2468\rangle, \langle 2478\rangle, \\
 \langle 2567\rangle, 
 \langle 2678\rangle, \langle 3457\rangle, \langle 3458\rangle, \langle 3468\rangle, 
 \langle 3478\rangle, \langle 3567\rangle, \langle 3678\rangle, \langle 4568\rangle, 
 \langle 4578\rangle, \langle 4678\rangle\}
\end{align*}
which can appear before $\langle1478\rangle$.

\bibliographystyle{utphys}
\bibliography{main}
\end{document}